\documentclass[twocolumn]{aastex7}

\shorttitle{prediction of the polar field from October 2025 to October 2026}
\shortauthors{Wang, Jiang, \&Luo}

\graphicspath{{./}{figures/}}

\usepackage{CJK}
\usepackage{multirow}
\usepackage{booktabs}
\usepackage{amsmath}
\usepackage{threeparttable}
\usepackage{verbatim}

\usepackage{hyperref}

\begin{document}
\begin{CJK*}{UTF8}{gbsn}

\title{Prediction of the Solar Polar Fields in 2026: An Unusually Weak Level Across the Last Five Solar Cycles}

\correspondingauthor{Jie Jiang}
\email{jiejiang@buaa.edu.cn}

\author[orcid=0000-0003-0256-8102,sname=Ruihui,gname=Wang]{Ruihui Wang (王瑞慧)}
\affiliation{School of Space and Earth Sciences, Beihang University, Beijing, People’s Republic of China}
\affiliation{Key Laboratory of Space Environment Monitoring and Information Processing of MIIT, Beijing, People’s Republic of China}
\email{wangruihui@buaa.edu.cn} 

\author[orcid=0000-0001-5002-0577,sname=Jie,gname=Jiang]{Jie Jiang (姜杰)}
\affiliation{School of Space and Earth Sciences, Beihang University, Beijing, People’s Republic of China}
\affiliation{Key Laboratory of Space Environment Monitoring and Information Processing of MIIT, Beijing, People’s Republic of China}
\email[show]{jiejiang@buaa.edu.cn}  

\author[orcid=0000-0002-6977-1239,sname=Yukun,gname=Luo]{Yukun Luo (罗昱琨)}
\affiliation{School of Space and Earth Sciences, Beihang University, Beijing, People’s Republic of China}
\affiliation{Key Laboratory of Space Environment Monitoring and Information Processing of MIIT, Beijing, People’s Republic of China}
\email{luoyukun@buaa.edu.cn}

\begin{abstract}

Solar polar fields are essential for the solar cycle and the heliospheric magnetic field. Cycle 25 is now entering its declining phase, the critical period during which most of the cycle's polar fields are established. Therefore, reliable polar-field prediction is now especially important. Polar-field evolution is governed by the poleward transport of already-emerged active-region (AR) flux over a timescale of a few years. Thus, surface flux-transport models can reliably provide one-year predictions without requiring information about future AR emergence. Our prediction method is validated using simulations of the surface magnetic field from 2020--2025 and hindcasts of the 2023--2024 polar fields, employing a newly constrained profile of the meridional flow. Using the most recent HMI synoptic magnetogram as the initial condition, we predict the polar-field evolution from October 2025 to October 2026. The southern polar field is predicted to strengthen gradually, while the northern field is expected to decline sharply until March 2026 due to some ARs with abnormal polarity. By that time, the northern polar field becomes exceptionally weak, and the southern field remains relatively weak, raising concerns about the polar-field strength at the cycle 25/26 minimum and the amplitude of cycle 26.

\end{abstract}


\section{Introduction} \label{sec:intro}

Solar polar fields affect the large-scale heliospheric magnetic field and the fast solar wind speed \citep{WangYM1990, Owens2013}, and they are especially important for the solar cycle evolution. Their polarity reversal marks the cycle maximum, and the polar-field strength at solar minimum provides a reliable precursor of the next cycle's amplitude \citep{Schatten1978, Jiang2007, Petrovay2020SCpd}. As cycle 25 enters its declining phase, when most of the cycle’s polar fields are established, accurately predicting their evolution becomes especially important.

The evolution of the polar fields is driven by the poleward transported flux of active regions (ARs). After emergence on the solar surface, AR flux is transported by differential rotation, meridional flow, and supergranular diffusion. A fraction of this flux is eventually transported to the poles, where it reverses the existing polar fields and builds new ones of opposite polarity \citep{Leighton1964, Wang1991}. This transport process is well described by surface flux transport (SFT) models \citep{WangYM1989ApJ, Jiang2014ssr, Yeates2023, Jha2024} and therefore, the polar fields can also be well predicted with SFT models \citep{Cameron2016, Hathaway2016, Jiang2018_apj}. However, SFT models still face two major limitations: the treatment of sources (ARs or sunspots) and the choice of flux-transport parameters. These limitations also constrain polar-field prediction. 

AR emergence involves significant stochasticity, and AR properties such as morphology and tilt angles also exhibit substantial variability. As a result, predicting future AR emergence is difficult and involves large uncertainties. Polar-field predictions that require generating future ARs—such as \cite{Cameron2016}, \cite{Jiang2018_apj}, and \cite{Jha2024}—therefore carry large uncertainties, with source-related uncertainty being the dominant contributor. However, for predicting the polar field only 1--2 years ahead, ARs that emerge during the prediction window are not crucial, because the transport of AR flux to the poles typically requires 1--2 years \citep{Jiang2014apj, Petrovay2020AM, 2020Yeates}. Thus, the SFT model combined with newly emerged ARs can be used for short-term (1--2 year) polar-field prediction. The most recent ARs can be assimilated accurately using the latest synoptic magnetograms as the initial field. Although this prediction method is conceptually straightforward, it has rarely been applied in polar-field prediction studies. One exception is \cite{Jiang2018_JASTP}, which postdicted the 2016--2017 polar fields and then predicted the 2017--2018 evolution, both with good agreement.

The flux-transport parameters, including the meridional flow and the supergranular diffusivity, govern the transport of AR flux and strongly influence the polar fields \citep{WangYM1989ApJ}. The supergranular diffusivity and the equatorial latitudinal gradient of the meridional flow control the evolution of the axial dipole strength \citep{Petrovay2020AM}, which serves as a proxy for the global polar fields. The latitude and magnitude of the peak flow speed, along with the flow profile from the peak latitude to the pole, further affect the strength and distribution of the polar fields, especially at solar minimum \citep{DeVore1984, WangYM2009, WangYM2017}. However, these flux-transport parameters remain poorly constrained by observations \citep{Jiang2023} and are usually determined by tuning the simulations to match the observations \citep{Lemerle2015, Whitbread2017}, most notably the axial dipole strength and the magnetic butterfly diagram. 

As a result, although many SFT studies successfully reproduce the axial dipole evolution, relatively few are able to reproduce the observed temporal evolution of the polar fields themselves. To address this limitation and, more importantly, to enhance the reliability of polar-field predictions, we derive a meridional-flow profile by optimizing the simulated polar fields against observations of the Helioseismic and Magnetic Imager on board the Solar Dynamics Observatory (SDO/HMI; \citealt{HMI}). Using this optimized meridional flow and the HMI synoptic magnetogram from Carrington rotation (CR) 2303 as the initial field, we predict the polar-field evolution from October 2025 to October 2026.

This paper is organized as follows. Section \ref{sec:model} describes the SFT model including the newly constrained meridional flow. Section \ref{sus:postdiction} validates the prediction method using this flow profile. Section \ref{sus:prediction} presents the predicted polar-field evolution from October 2025 to October 2026. Section \ref{sus:comparision_cycles_2125} compares the cycle-25 polar fields with those of earlier cycles. Section \ref{sec:conclusion} summarizes the main results.

% One is to apply the statistical properties of former emerged ARs (sunspot groups) and run Monte Carlo ensembles to generate ARs \citep{Cameron2016, Jiang2018_apj, Bhowmik2018}. The other is use a past cycle to act as a proxy for AR emergence in the predicted cycle, such as choosing cycle 14 as a proxy of cycle 24 \citep{Upton2014, Hathaway2016, Upton2018, Jha2024}. However, the prediction of AR emergence is difficult and the above two methods both introduce large uncertainty in the polar fields prediction.

\section{Model Description}\label{sec:model}

The surface flux transport (SFT) model is governed by the following equation: 
\begin{equation}
\label{eq:SFT}
\begin{split}
\frac{\partial B}{\partial t} &= -\omega(\theta) \frac{\partial B}{\partial \phi} - \frac{1}{R_\odot \sin \theta} \frac{\partial}{\partial \theta} [u(\theta) B \sin \theta] \\
&+ \frac{\eta}{R_\odot^2} \left[ \frac{1}{\sin \theta} \frac{\partial}{\partial \theta} \left( \sin \theta \frac{\partial B}{\partial \theta} \right) + \frac{1}{\sin^2 \theta} \frac{\partial^2 B}{\partial \phi^2} \right]\\
&+ S(\theta, \phi, t),
\end{split}
\end{equation}
where $B$ is the radial magnetic field, and $\theta$ and $\phi$ represent colatitude and heliographic longitude, respectively. The differential rotation $\omega(\theta)$ follows the profiles from \cite{Snodgrass1983}, and $\eta$ is the supergranular diffusivity. The meridional flow $u(\theta)$ remains poorly constrained by observations. Although several commonly used profiles exist \citep{van_Ballegooijen1998, Lemerle2017, WangYM2017, Whitbread2018}, they mainly show good performance in reproducing the observed axial dipole strength and magnetic butterfly diagram. To better simulate the polar fields, we derive a meridional flow profile by optimizing the simulated mean fields against HMI observations within $60^\circ$--$75^\circ$ latitude. This profile, denoted as Flow 1, is constructed based on the functional form proposed by \citet{Lemerle2015} and is expressed as:
\begin{equation}
\label{eq:MF_new}
u(\theta) = u_0 \operatorname{erf}^8(2.25 \sin \theta) \operatorname{erf}(3 \cos \theta),
\end{equation}
where $u_0 = 13 m/s$. 

\begin{figure}[htbp!]
\centering
\includegraphics[scale=0.33]{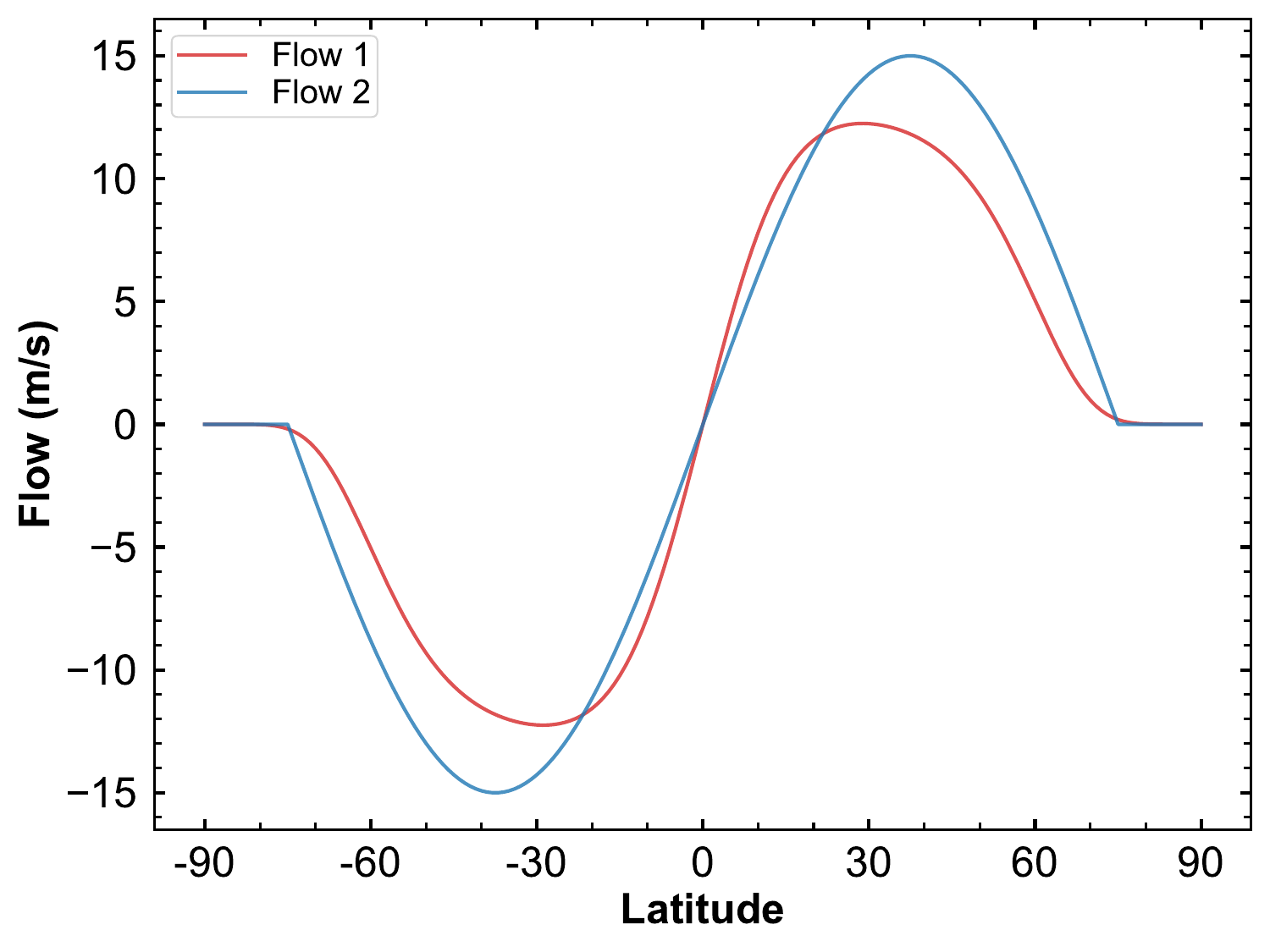}
\caption{Comparison of the two meridional flows. Flow 1 is the profile obtained by optimizing the simulated polar fields within $60^{\circ}$--$75^{\circ}$ latitude against HMI observations. Flow 2 follows the form given by \cite{van_Ballegooijen1998}.
\label{fig:MF}}
\end{figure}

To evaluate the performance of Flow 1, we also implement the widely used meridional flow profile from \citet{van_Ballegooijen1998}, denoted as Flow 2, defined as:
\begin{equation}
\label{eq:MF_VB}
u(\lambda) = 
\begin{cases}
u_0 \sin \left( \pi \lambda / \lambda_0 \right) & \text{if } |\lambda| < \lambda_0 \\
0 & \text{otherwise},
\end{cases}
\end{equation}
where $\lambda$ is latitude, and $u_0 = 15 m/s$. The shapes of both flow profiles are illustrated in Figure \ref{fig:MF}. Compared to Flow 2, Flow 1 is slower between roughly $20 ^{\circ}$ and $70 ^{\circ}$ latitude. Its peak speed is both lower in magnitude and shifted equatorward, occurring at $29^{\circ}$. The equatorial latitudinal gradient ($D_u$) of Flow 1 is about $0.9~m\,s^{-1}\,{deg}^{-1}$ for $u_0 = 13$ m/s, which is consistent with most helioseismology measurements \citep{Zhao2014, Gizon2020, Jiang2023} and steeper than that of Flow 2. A comparison of their simulation results is presented in Section \ref{sus:postdiction}. Further discussion of Flow 1 will be provided in future work.

The supergranular diffusivity and the equatorial latitudinal gradient of the meridional flow jointly determine the evolution of the axial dipole strength of ARs and the axial dipole strength of the global surface field \citep{Petrovay2020AM}. To enable a fair comparison, we adjust the diffusivity for each flow to yield identical global axial dipole strengths: $\eta= 450 km^2/s$ for Flow 1 and $300 km^2/s$ for Flow 2.

\begin{figure*}[htbp]
\centering
\includegraphics[scale=0.33]{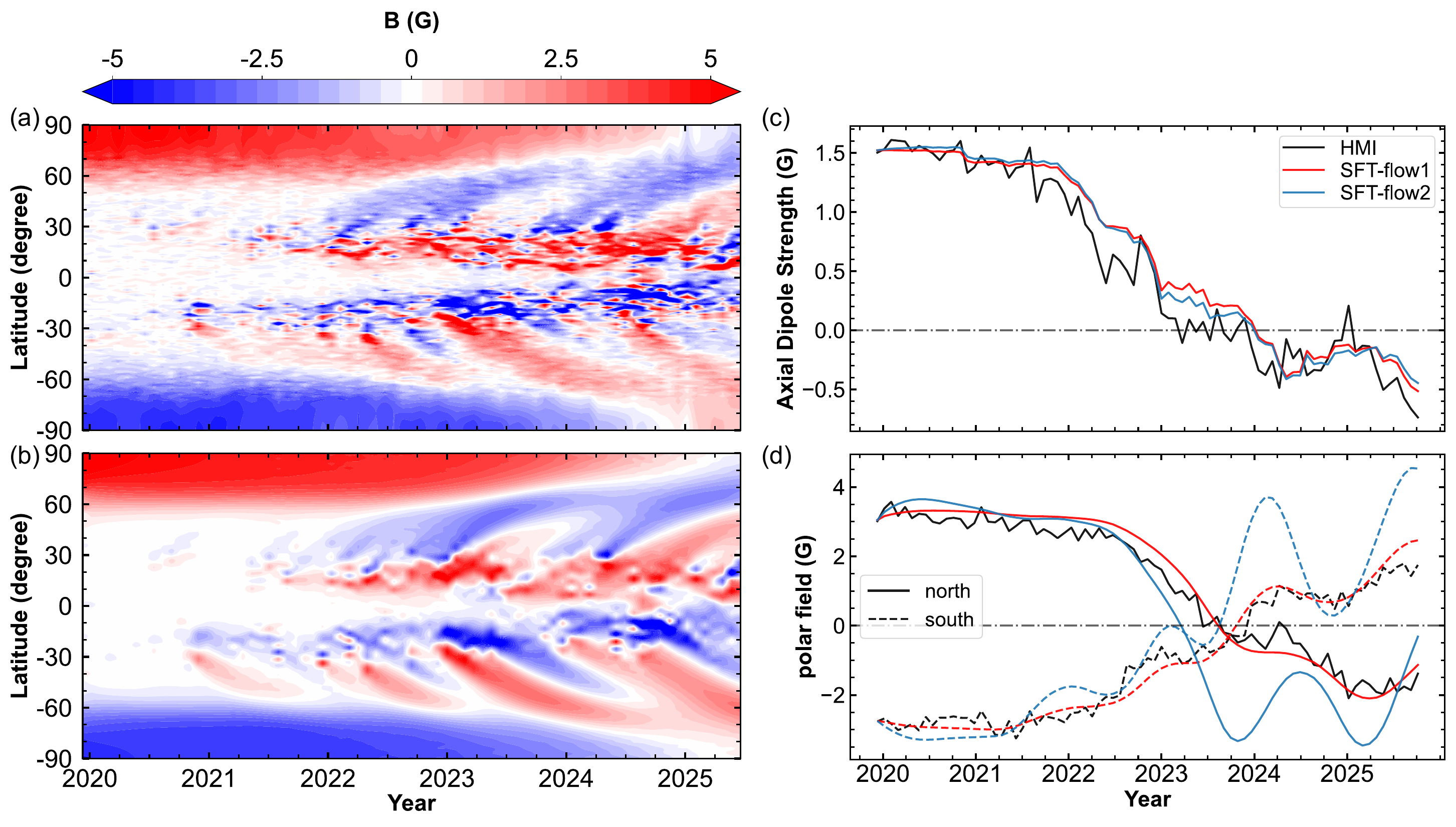}
\caption{Comparison of the surface magnetic field from HMI synoptic magnetograms and the SFT simulation during 2020--2025.
Panels (a) and (b) show the magnetic butterfly diagrams from the HMI observations and the SFT simulation with Flow 1, respectively. Panel (c) presents the axial dipole strength, and Panel (d) shows the polar fields, averaged over $60^\circ$-$75^\circ$ latitude. The dash-dotted lines in Panels (c) and (d) indicate the zero level. Panels (c) and (d) each show two SFT simulations using Flow 1 and Flow 2, whose profiles are given in Equations \ref{eq:MF_new} and \ref{eq:MF_VB}.
\label{fig:sml_hmi}}
\end{figure*}

The source term $S(\theta, \phi, t)$ is obtained by assimilating ARs from the Active Region database for Influence on Solar cycle Evolution (ARISE; \citealt{Wang2023, Wang2024}). It is used only when validating the prediction method (Section \ref{sus:postdiction}). For short-term polar-field predictions, the source is not required, and we set $S(\theta,\phi,t)=0$. We solve the SFT equation using our newly developed SFT code based on spectral methods, which accurately reproduces the butterfly diagram, axial dipole strength, and observed magnetic power spectra \citep{wang2025, Luo2025}.

\section{results}

% \subsection{Forecasting the polar field in about the future one year: June 2025-June 2026 }\label{sec: prediction}

\begin{figure*}[htbp]
\centering
\includegraphics[scale=0.4]{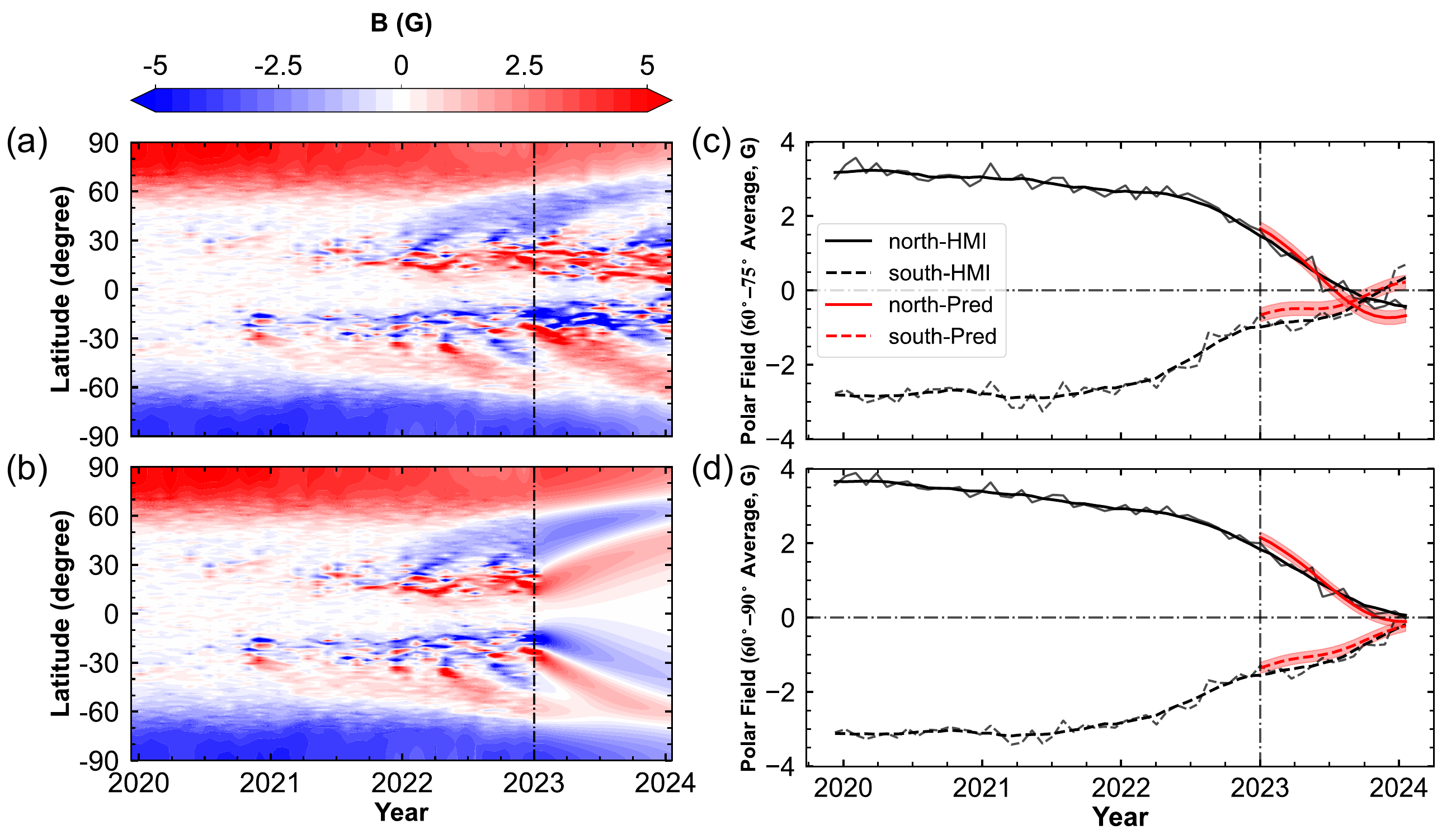}
\caption{
Comparison of magnetic butterfly diagrams and polar fields between HMI observations and SFT predictions for 2023--2024.
(a) HMI magnetic butterfly diagram.
(b) Magnetic butterfly diagram combining HMI observations (before 2023) and SFT predictions (from 2023 onward).
(c) Comparison of polar fields averaged over $60^{\circ}-75^{\circ}$.
(d) Comparison of polar fields averaged over $60^{\circ}-90^{\circ}$.
The vertical dash-dotted line marks the start of the prediction, and the horizontal dash-dotted line marks the zero level.
In (c) and (d), solid and dashed lines show the northern and southern polar fields. Thick black lines show the nine-CR moving-averaged fields, thin gray lines the original fields, and red lines the predicted fields, with red shading indicating the prediction uncertainties.
\label{fig:test_pred}}
\end{figure*}

\subsection{Validation of the Prediction Method} \label{sus:postdiction}

Before applying the SFT model with the constrained meridional flow to predict the polar fields for the coming year, we first simulate the surface magnetic field of the ongoing cycle 25 and perform a post-prediction of the polar fields for 2023--2024 to further assess the model performance and the reliability of the prediction method.

Using the HMI synoptic magnetogram of CR 2225 as the initial condition, we simulate the evolution of the solar surface magnetic field from CR 2225 to CR 2303 (December 2019--October 2025). Figure \ref{fig:sml_hmi} presents the results from simulations with Flow 1 and Flow 2, together with those from pole-filled HMI synoptic maps, including the magnetic butterfly diagram, the axial dipole strength, and the polar fields averaged over $60^\circ$--$75^\circ$. The transport parameters in both simulations are adjusted to yield the same axial dipole strength (Figure \ref{fig:sml_hmi} (c)). The constrained flow (Flow 1) reproduces all three observational features, demonstrating its good performance. Flow 2, although capable of reproducing the axial dipole strength and the magnetic butterfly diagram (not shown), fails to match the observed polar fields (Figure \ref{fig:sml_hmi} (d)), with large deviations emerging from 2023 onward, around and after the time of the polarity reversal. 

We then hindcast the polar fields during 2023--2024, a period characterized by prominent surges. Because the polar-field evolution is strongly modulated by these surges, predictions during 2023--2024 are particularly challenging and therefore provide a stringent test of the model accuracy. Using the HMI synoptic magnetogram of CR 2266 as the initial condition, we simulate the surface magnetic field evolution over 2023--2024, with results shown in Figure \ref{fig:test_pred}. Figure \ref{fig:test_pred}(a) displays a prominent negative surge in the northern hemisphere and a prominent positive surge in the southern hemisphere during 2023--2024. The simulated butterfly diagram in Figure \ref{fig:test_pred}(b) reproduces these surges, although the simulated southern surge captures only part of the observed one. This is because the observed southern surge results from a sequence of active regions emerging over roughly half a year, many of which appeared after CR 2266 and thus are not included in the simulation. The simulation also produces a positive surge in the north and a negative surge in the south, originating from the leading-polarity flux of ARs emerging during and slightly before CR 2266. These features do not appear in the observations because they were later declined by ARs that emerged after CR 2266.

Figures \ref{fig:test_pred} (c) and (d) show the predicted northern and southern polar fields, together with their associated uncertainties for 2023--2024. The polar field is computed as
\begin{equation}
\label{eq:polar_field}
B_f
=
\frac{
\displaystyle
\sum_{\lambda_i=\lambda_1}^{\lambda_2}
\sum_{\phi_j=0}^{2\pi}
B_{i,j}\,\cos\lambda_i\,\Delta\lambda\,\Delta\phi
}{
\displaystyle
\sum_{\lambda_i=\lambda_1}^{\lambda_2}
\sum_{\phi_j=0}^{2\pi}
\cos\lambda_i\,\Delta\lambda\,\Delta\phi
},
\end{equation}
where $B_{i,j}$ denotes the magnetic field strength at the grid point $(i,j)$, and $\Delta\lambda$ and $\Delta\phi$ are the latitudinal and longitudinal grid spacings, respectively. The parameters $\lambda_1$ and $\lambda_2$ define the latitudinal range over which the polar field is averaged. In this study, the ranges $60^{\circ}$--$75^{\circ}$ and $60^{\circ}$--$90^{\circ}$ are adopted, corresponding to Figures \ref{fig:test_pred} (c) and (d), respectively. Equation \ref{eq:polar_field} is also equivalent to the arithmetic mean in the equal-sine-latitude coordinates used by the HMI synoptic magnetograms.

The total prediction uncertainty has two components. The first contribution is the HMI measurement limitation ($\sigma_{HMI}$), estimated from the mean net magnetic field in the HMI synoptic maps over the three CRs nearest to the target time. For an ideal measurement, the net flux should be zero; the non-zero values reflect instrumental limitations. This approach follows \citet{Cameron2016}. The resulting $\sigma_{HMI}$ is about 0.059 G. 

The second contribution comes from fluctuations of the observed polar fields around their time-averaged values ($\sigma_{smt}$). As shown in Figure \ref{fig:sml_hmi}, the simulated polar fields are smoother than the HMI observations, indicating that the SFT model reproduces only the time-mean polar field. These short-term observational fluctuations therefore introduce discrepancies between the observed and predicted polar fields.
These deviations, computed using the combined northern and southern polar fields, are about 0.175 G for the $60^{\circ}$--$75^{\circ}$ averages and about 0.136 G for the $60^{\circ}$--$90^{\circ}$ averages. They are the dominant source of uncertainty. The total prediction error is the quadratic sum of the two terms, $\sigma= \sqrt{\sigma_{HMI}^2+\sigma_{smt}^2}$， yielding uncertainties of about 0.185 G for the $60^{\circ}$--$75^{\circ}$ averages and 0.148 G for the $60^{\circ}$--$90^{\circ}$ averages.

As shown in Figure \ref{fig:test_pred}, the predicted mean fields at $60^{\circ}$--$75^{\circ}$ and $60^{\circ}$--$90^{\circ}$ both agree well with the observations, and the observed values fall within or close to the predicted uncertainties. Because the $60^{\circ}$--$90^{\circ}$ range includes stronger high-latitude fields, its mean values are larger than those at $60^{\circ}$--$75^{\circ}$ during 2020--2023, and its polarity reversal occurs later.

However, small deviations remain between the predicted and observed polar fields, particularly for the $60^{\circ}$–-$75^{\circ}$ averages. The deviation at $60^{\circ}$–-$90^{\circ}$ is smaller because the flux from ARs emerging during 2023--2024 cannot reach latitudes above $75^{\circ}$ by the end of 2024. For the $60^{\circ}$--$75^{\circ}$ averages, the deviation gradually increases with time. This is likely due to the absence of far-side ARs in the initial magnetograms: the most recently emerged far-side ARs are missing from the synoptic map of CR 2266 and gradually affect the polar fields as they evolve through 2023--2024, although the effect remains limited, as shown in Figure \ref{fig:test_pred} (c).

In addition, the transport parameters used in the simulation strongly affect the prediction. These parameters are typically optimized through comparison between simulations and observations. In the post-prediction test, the optimization period includes the prediction interval, so the resulting parameters remain appropriate. In real predictions, however, the parameters may no longer be optimal and can become a significant source of deviation between the predicted and observed polar fields.

These limitations also equally apply to SFT predictions that generate future ARs. Future work will address these issues or explicitly quantify their contribution to the prediction errors.

\subsection{Prediction of the Polar Fields from October 2025 to October 2026} \label{sus:prediction}

\begin{figure}[htbp!]
\centering
\includegraphics[scale=0.33]{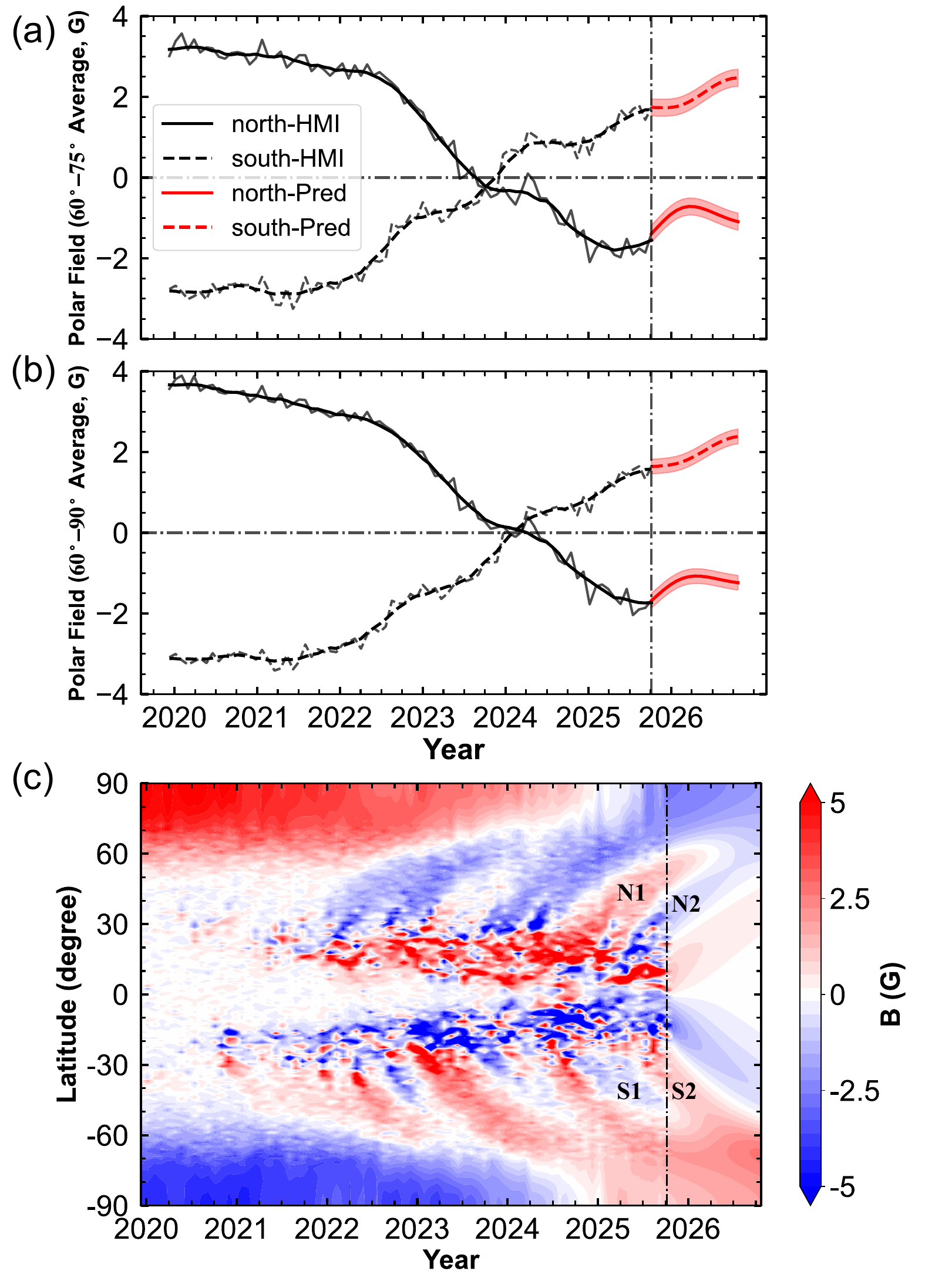}
\caption{Prediction of polar fields averaged over $60^{\circ}-75^{\circ}$ (a), polar fields averaged over $60^{\circ}-90^{\circ}$ (b), and the magnetic butterfly diagram (c) for CRs 2303--2317 (2025.10--2026.10). Line styles and colors follow the same convention as in Figure \ref{fig:test_pred}.
\label{fig:predt}}
\end{figure}

After validating the reliability of the prediction method, we predict the polar field from October 2025 to October 2026 using the HMI synoptic map of CR 2303 as the initial field. The results are shown in Figure \ref{fig:predt}, including the polar fields averaged over $60^{\circ}$--$90^{\circ}$, the polar fields averaged over $60^{\circ}$--$75^{\circ}$, and the corresponding magnetic butterfly diagram.

Panels (a) and (b) show that the northern polar field will experience a pronounced decline until about March 2026, followed by a gradual recovery. The decrease is weaker in the $60^{\circ}$--$90^{\circ}$ averages than in the $60^{\circ}$--$75^{\circ}$ averages. This behavior is explained by the predicted magnetic butterfly diagram in panel (c). A strong positive surge (N1) appears near $50^{\circ}$ north before March 2026, which significantly weakens the northern polar field. This surge directly continues the one observed by HMI from June 2024 to September 2025 and is produced by ARs, including some with abnormal polarity, that emerged between roughly June 2024 and March 2025. Because the flux from ARs emerging during the prediction interval is confined to lower latitudes, it has little influence on this high-latitude surge. As a result, the decline in the northern polar field is unavoidable.

The ARs that emerged in the latest few CRs, including CR 2303, generate a negative surge (N2). After the surge N1 dissipates around March 2026, the negative surge N2 begins to strengthen the northern polar field.

As for the southern polar field, panels (a) and (b) show that it remains nearly unchanged during the first three months and then gradually increases. The following polarities of ARs emerging from March 2025 to September 2025 generate a positive surge (S2), which persistently influences the polar region throughout the prediction interval. At the beginning of the interval, this positive surge is partly canceled by an existing weak negative surge (S1) at high latitudes, keeping the polar field stable. After the negative surge fully disappears, S2 continues to strengthen the southern polar field.

Two limitations in the prediction method, discussed in Section \ref{sus:postdiction}, may introduce differences between the simulated flux transport and future HMI observations. These limitations are the absence of some far-side ARs in the initial maps and the choice of transport parameters. Although the positive surge N1, which drives the decline of the northern polar field, is unaffected by the possible missing ARs in the initial map, its migration is influenced by the transport parameters. Differences in this migration can cause the predicted decrease in the northern polar field to occur earlier or later than observed.

\subsection{Phase-Aligned Comparison of Polar Fields Across the Last five Cycles} \label{sus:comparision_cycles_2125}

\begin{figure*}[htbp!]
\centering
\includegraphics[scale=0.33]{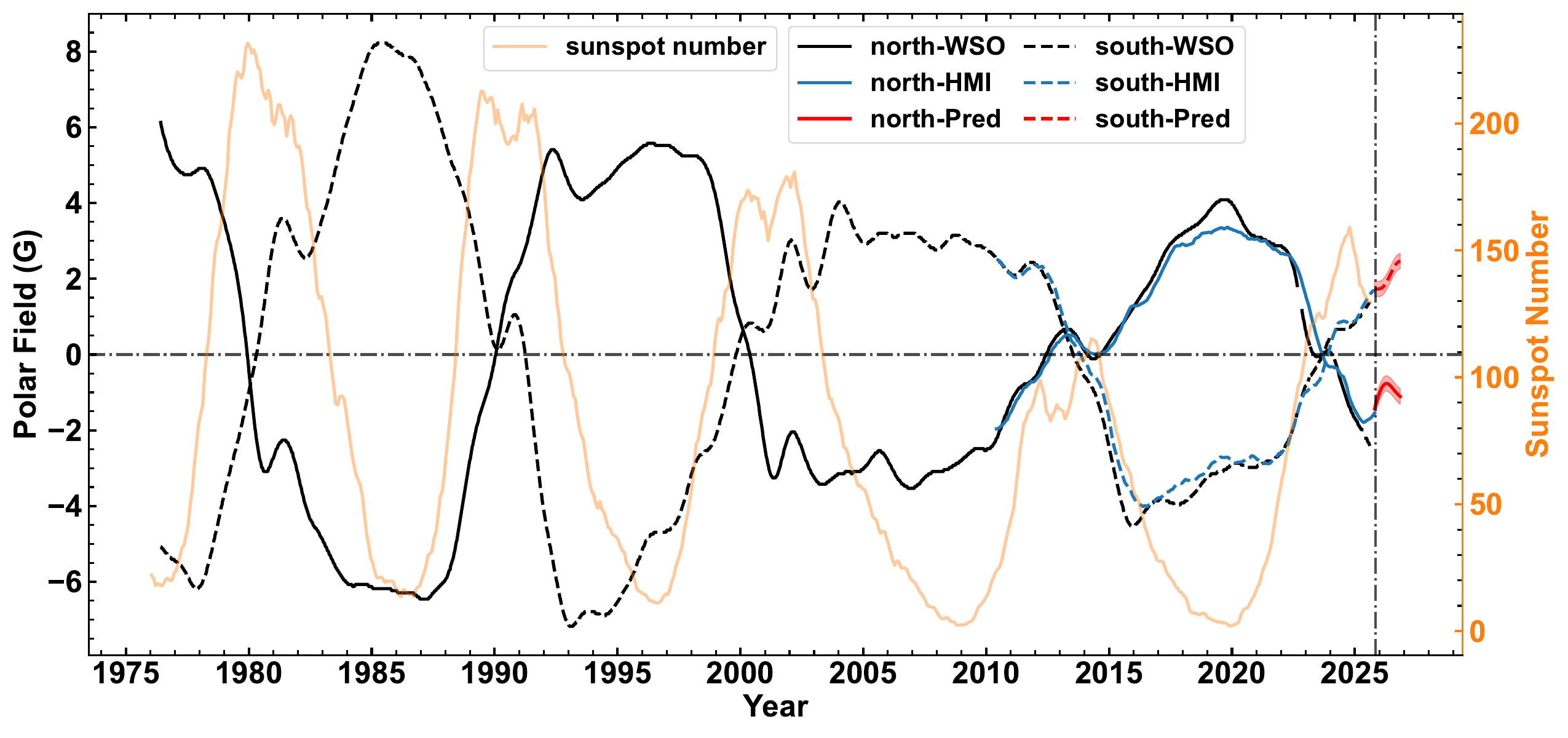}
\caption{Comparison of the predicted unusually weak polar fields across the five cycles. The WSO polar fields (black) are averaged over latitudes $55^\circ$--$90^\circ$, while the HMI fields (blue) and SFT predictions (red) are averaged over latitudes $60^\circ$–-$75^\circ$. The WSO line-of-sight magnetic fields are converted to radial fields by dividing by a projection factor of 0.2482 and then calibrated to the HMI fields by multiplying by 1.37. The orange curve shows the sunspot number from WDC-SILSO for reference. The vertical dash-dotted line indicates the start of the prediction.
\label{fig:wso-1y}}
\end{figure*}

\begin{table}[htbp]
\centering
\caption{Comparison of WSO and predicted polar fields for solar cycles 21--25}
\label{tab:polar_field}
\begin{tabular}{ccccc}
\hline \hline
Cycle &       & $B_{t=6y}(G)$ \textsuperscript{a} & $B_{t=min}(G)$ \textsuperscript{b} & $B_{t=6y}/B_{t=min}$ \\
\hline
      & North & -4.35       & -6.30                  & 69.1\% \\
\multirow{-2}{*}{21} & South & 2.68        & 7.81                   & 34.3\% \\
\hline
      & North & 4.56        & 5.53                   & 82.5\% \\
\multirow{-2}{*}{22} & South & -7.11       & -4.69                  & 151.6\% \\
\hline
      & North & -3.14       & -2.83                  & 110.7\% \\
\multirow{-2}{*}{23} & South & 1.72        & 3.12                   & 55.2\% \\
\hline
      & North & 0.56        & 4.05                   & 13.9\% \\
\multirow{-2}{*}{24} & South & -3.19       & -2.95                  & 108.1\% \\
\hline
      & North & $-0.72\pm 0.21$   &                        &        \\
\multirow{-2}{*}{25} & South & $1.96\pm 0.21$        &                        &        \\
\hline
\end{tabular}

\smallskip
\raggedright
\textsuperscript{a} Polar field six years after cycle start; values for cycle 25 are predicted and averaged over $60^{\circ}$--$75^{\circ}$. \\
\textsuperscript{b} Polar field at the solar minimum.
\end{table}

Our prediction shows that by March 2026, the northern polar field averaged over $60^{\circ}-75^{\circ}$ remains weak at only $-0.72 \pm0.21$ G, even though six years have passed since the start of cycle 25. To compare the polar fields with those of earlier cycles, Figure \ref{fig:wso-1y} presents Wilcox Solar Observatory (WSO) measurements for cycle 25 alongside cycles 21--24. The WSO line-of-sight magnetic fields are converted to radial components following \cite{Hathaway2016}, where the original measurements are divided by 0.2482, the average projection factor for latitudes between $55^{\circ}$ and $90^{\circ}$. The resulting radial fields are then calibrated to match HMI mean magnetic fields over $60^{\circ}-75^{\circ}$ by applying a scaling factor of 1.37. The calibrated WSO radial fields agree well with HMI data, especially in the timing of polar field reversals. In contrast, the agreement is poorer when comparing WSO radial fields with HMI mean fields over $60^{\circ}-90^{\circ}$, particularly in reversal timing, so those comparisons are not shown. To help the comparisons across cycles, we list the polar fields six years after the start of each cycle ($B_{t=6y}$) in Table \ref{tab:polar_field}.

The WSO polar fields show that such a weak $B_{t=6y}$ as in cycle 25 is uncommon among cycles 21--25. In the north, the cycle-25 value of $-0.72 \pm 0.21$ G is far smaller than those of cycles 21--23, all of which exceed 3 G. Although it is comparable to the northern $B_{t=6y}$ of cycle 24, the southern value in cycle 25 is much weaker than that of cycle 24. The weak polar fields six years into cycle 25, especially in the north, raise concerns about the polar field strength at the upcoming solar minimum and therefore the amplitude of cycle 26. 

However, the polar field strength at the solar minimum ($B_{t=\min}$) between cycles 25 and 26 remains highly uncertain. As shown in Table \ref{tab:polar_field}, the ratio $B_{t=6y}/B_{t=\min}$ varies widely from cycle to cycle, ranging from 13.9\% to 151.6\%. This large spread indicates that the polar field six years into a cycle does not uniquely determine the polar field strength at the subsequent minimum. In particular, for the northern hemisphere of cycle 24, $B_{t=6y}$ accounts for only 13.9\% of $B_{t=\min}$, implying that the majority of the polar field was built up during the declining phase of that cycle, as also illustrated in Figure \ref{fig:wso-1y}. Consequently, if a similar buildup during the declining phase of cycle 25 takes place, the polar fields at the cycle-25/26 minimum could be comparable to, or even stronger than, those at the cycle-24/25 minimum.

\section{Conclusion} \label{sec:conclusion}

In this paper, we use an SFT model without a source term in the flux transport equation to forecast the polar fields one year ahead. To better reproduce the observed polar fields, we apply a meridional flow obtained by optimizing the simulated polar fields to match the HMI observations. This flow successfully reproduces the magnetic butterfly diagram, axial dipole strengths, and, in particular, the polar fields from 2020 to 2025. Predictions using this flow also accurately hindcast the observed polar fields at latitudes $60^{\circ}-75^{\circ}$ and $60^{\circ}-90^{\circ}$ during 2023--2024. These results demonstrate the feasibility of short-term polar-field prediction that does not rely on future AR emergence, and highlight the clear advantage of the optimized meridional-flow profile in reproducing and predicting polar-field evolution.

The prediction for October 2025 to October 2026 shows a gradual strengthening of the southern polar field, while the northern polar field undergoes a sharp decline until March 2026. This decline is driven by a pre-existing positive magnetic surge, which is probably caused by some ARs with abnormal polarity. Although uncertainties in the transport parameters may slightly affect the migration of this surge relative to observations, the decline itself is unavoidable because the surge originates from earlier ARs and is only weakly affected by ARs emerging during the prediction interval. As a result, observations around March 2026 are expected to show an unusually weak northern polar field, even as cycle 25 enters its declining phase. Compared with previous cycles, the mean polar field of cycle 25 is the weakest at a similar phase. This raises concerns about the polar-field strength at the upcoming solar minimum and the amplitude of cycle 26, although the polar fields could quickly strengthen in the following years.

As the Solar Orbiter can now observe the polar magnetic field from outside the ecliptic plane, the quality of polar-field measurements will be improved \citep{Calchetti2025}. This advancement will also help enhance future simulations and predictions of the polar field.

% Without generating new ARs, our method yields smaller prediction errors than earlier approaches \citep{Cameron2016, Hathaway2016, Jha2024}. The remaining uncertainties arise from measurement errors in the HMI synoptic maps, short-term polar-field fluctuations, the absence of far-side ARs in the initial field, and uncertainties in the transport parameters. The measurement errors and polar-field fluctuations are already included in our error estimates. These four limitations equally apply to SFT predictions that generate future ARs. Future work should address these issues or explicitly quantify their contribution to the prediction errors. Finally, this method cannot predict the future axial dipole strength, which changes immediately once new ARs emerge.

%\\ \hspace*{\fill} \\

\begin{acknowledgements}
The research is supported by the National Natural Science Foundation of China (grant Nos. 12425305, 12350004, and 12173005) and China's Space Origins Exploration Program. The SDO/HMI data are courtesy of NASA and the SDO/HMI team. Wilcox Solar Observatory data used in this study were obtained via the website http://wso.stanford.edu, courtesy of J.T. Hoeksema. The sunspot number is from WDC-SILSO, Royal Observatory of Belgium, Brussels, DOI: https://doi.org/10.24414/qnza-ac80.
\end{acknowledgements}

\bibliography{sample7}{}
\bibliographystyle{aasjournal}

% \appendix  

% Figure \ref{fig:MF} compares the two meridional flows used in the validation of the new flow and prediction method, Section \ref{sus:postdiction}. The exact forms of Flow 1 and Flow 2 are given in Equations \ref{eq:MF_new} and \ref{eq:MF_VB}, respectively.

\end{CJK*}
\end{document}